\documentclass[12pt,english]{iopart}
\usepackage{cite}
\usepackage{graphicx}
\usepackage{iopams}
\usepackage{bm}
\usepackage[usenames,dvipsnames,svgnames,table]{xcolor}
\eqnobysec
\usepackage{amssymb}
\usepackage{graphicx}
\usepackage{mathrsfs}
\usepackage{ntheorem}
\usepackage{enumerate}
\usepackage{enumitem}
\usepackage{times,txfonts}
\usepackage{subfigure}
\newcommand{\ket}[1]{|#1\rangle}
\newcommand{\bra}[1]{\langle #1|}
\usepackage[title,titletoc,header]{appendix}

\newcommand{\I}{\mathrm{I}}

\newcommand\abs[1]{\left|#1\right|}

\def\CC{{\rm\kern.24em \vrule width.04em height1.46ex depth-.07ex \kern-.30em C}}
\def\RR{{\rm\kern.24em \vrule width.04em height1.46ex depth-.07ex
\kern-.30em R}}
\def\P{{\rm I\kern-.25em P}}
\usepackage[unicode=true,
            pdfusetitle,
            bookmarks=true,
            bookmarksnumbered=false,
            bookmarksopen=false,
            breaklinks=true,
            pdfborder={0 0 0},
            backref=false,
            colorlinks=true]{hyperref}
\hypersetup{linkcolor=NavyBlue,urlcolor=NavyBlue,citecolor=NavyBlue}
\providecommand{\openone}{\leavevmode\hbox{\small1\kern-4.3pt
\normalsize1}}

\begin{document}

\title{Superadditivity of convex roof coherence measures}

\author{C. L. Liu}
\address{Department of Physics, Shandong University, Jinan 250100, China}

\author{Qi-Ming Ding}
\address{Department of Physics, Shandong University, Jinan 250100, China}

\author{D. M. Tong}
\address{Department of Physics, Shandong University, Jinan 250100, China}
\ead{tdm@sdu.edu.cn}

\begin{abstract}
In this paper, we examine the superadditivity of convex roof coherence measures. We put forward a theorem on the superadditivity of convex roof coherence measures, which provides a sufficient condition to identify the convex roof coherence measures fulfilling the superadditivity. By applying the theorem to each of the known convex roof coherence measures, we prove that the coherence of formation and the coherence concurrence are superadditive, while the geometric measure of coherence, the convex roof coherence measure based on linear entropy, the convex roof coherence measure based on fidelity, and convex roof coherence measure based on $\frac{1}{2}$-entropy are non-superadditive.
\end{abstract}
~~~~~~~~\providecommand{\keywords}[1]{Keywords:~~#1}
~~~\keywords{convex roof, coherence measures, superadditivity}

\maketitle

\section{Introduction}
Quantum coherence is an essential feature of quantum mechanics which is responsible for the departure between the classical and quantum world. It is an important component in quantum information processing \cite{Nielsen}, and plays a central role in emergent fields, such as quantum metrology \cite{Giovannetti,Giovannetti1}, nanoscale thermodynamics \cite{Aberg,Lostaglio,Lostaglio1}, and quantum biology \cite{Sarovar,Lloyd,Huelga,Lambert}.  Recently, the quantification of coherence has attracted a growing interest due to the development of quantum information science \cite{Aberg1,Bagan,Baumgratz,Bera,Bromley,Cheng,Chitambar,Chitambar1,Chitambar2,Chitambar3,Du,Du1,Wang,Bu,Girolami,Levi,Liu,Mani,Napoli,Qi,Peng,Yuchang,Liu2,Zhu,Liu1,
Radhakrishnan,Shao,Singh,Streltsov,Vicente,Winter,Xi,Yadin,Yao,Yu,Guo,Yuan,Zhang,Ma,Tan,Hu,Streltsov1}.

By adopting the viewpoint of coherence as a physical resource, Baumgratz \textit{et al.} proposed a seminal framework for quantifying coherence \cite{Baumgratz}.  In that framework, a functional of states can be taken as a coherence measure if it fulfills four conditions, namely, the coherence being zero (positive) for incoherent states (all other states), the monotonicity of coherence under incoherent operations, the monotonicity of coherence under selective measurements on average, and the nonincreasing of coherence under mixing of quantum states. By following the framework, a number of coherence measures have been found. Some of them are defined based on the distance between the state under consideration to the set of incoherent states \cite{Baumgratz,Shao,Radhakrishnan,Napoli,Streltsov}, such as the $l_1$ norm of coherence \cite{Baumgratz}, the relative entropy of coherence \cite{Baumgratz} and the robustness of coherence \cite{Napoli}, while others are defined based on the convex roof construction \cite{Zhu,Yuan,Du1,Liu1,Qi,Aberg1,Streltsov}, such as the coherence of formation \cite{Winter,Yuan,Aberg1}, the geometric measure of coherence \cite{Streltsov}, and the coherence concurrence \cite{Qi}, where the coherence of a mixed state is quantified by the weighted sum of the coherence of the pure states in a decomposition of the mixed state, minimized over all possible decompositions. With these coherence measures, various topics of quantum coherence, such as the dynamics of coherence \cite{Bromley,Yu}, the distillation of coherence \cite{Yuan,Winter, Liu2}, and the relations between quantum coherence and quantum correlations \cite{Streltsov,Xi,Yao,Ma,Tan,Yadin,Guo,Radhakrishnan} have been investigated.

Another interesting topic of quantum coherence is the superadditivity of a coherence measure. A coherence measure $C$ is said to be superadditive if the relation,
\begin{equation}
 C(\rho_{AB})\geq C(\rho_A)+C(\rho_B),
\label{CAB}
\end{equation}
is valid for all density matrices $\rho_{AB}$ of a finite-dimensional system with respect to a particular reference basis $\{\ket{i}_A\otimes \ket{j}_B \}$, where $\rho_A=\tr_B\rho_{AB}$ and $\rho_B=\tr_A\rho_{AB}$ are with respect to the basis  $\{\ket{i}_A\}$ and $\{\ket{j}_B\}$, respectively. The superadditivity of a coherence measure describes the trade-off relations between the coherence of a bipartite system and that of its subsystems and it is a precondition of defining a discordlike correlation based on the coherence measure \cite{Guo,Tan}.  Investigations on this topic have been started recently \cite{Guo,Tan,Xi,Bu,Wang}. The superadditivity of the relative entropy of coherence was first proved  in Ref. \cite{Xi}, and based on the superadditivity of the relative entropy of coherence, the discordlike correlations were established \cite{Guo,Wang}. The superadditivity of the $l_1$ norm of coherence was then proved in  Ref. \cite{Tan}, and based on it a correlated coherence describing the relationship between bipartite coherence and quantum correlations is defined. It was recently proved that the robustness of coherence is non-superadditive, i.e., not satisfying the superadditivity \cite{Bu}. Therefore, the superadditivity or non-superadditivity of all the known three coherence measures defined based on distance have been resolved. However, the superadditivity of convex roof coherence measures remains unresolved. Since convex roof coherence measures involve an optimization process, they usually do not admit a closed form expression for mixed states although they typically admit a closed form expression for pure states. Thus, it is more difficult in general to prove whether the superadditivity is valid for a convex roof coherence measure than that for a distance-based coherence measure.

In this paper, we address the issue: which of the known convex roof coherence measures are superadditive and which are non-superadditive?  To examine the superadditivity of a convex roof coherence measure, we will put forward a theorem, which provides a sufficient condition to identify the convex roof coherence measures fulfilling the superadditivity. By applying the theorem to each of the  known convex roof coherence measures, we find that the coherence of formation and the coherence concurrence are superadditive, while the geometric measure of coherence, the convex roof coherence measure based on linear entropy, the convex roof coherence measure based on fidelity, and convex roof coherence measure based on $\frac{1}{2}$-entropy are non-superadditive.

\section{\textbf{Convex roof coherence measures}}
To present our findings clearly, we first recapitulate some notions related to our topic.  Coherence of a state is measured with respect to a particular reference basis, whose choice is dictated by the physical scenario under consideration. If the particular basis is denoted as $\{\ket{i},~i=1,2,\cdot\cdot\cdot,d\}$, an incoherent state is then defined as $\delta=\sum_ip_i\ket{i}\bra{i}$, where $p_i$ are probabilities with $\sum_ip_i=1$. The set of all incoherent states is denoted by $\mathcal{I}$. All other states which cannot be written as diagonal matrices in this basis are called coherent states. We use $\rho$ to represent a general state, and $\delta$ specially to denote an incoherent state. An incoherent operation is defined as a completely positive trace-preserving map, $\Lambda(\rho)=\sum_nK_n\rho K_n^\dagger$,
where the Kraus operators $K_n$ satisfy not only $\sum_n K_n^\dagger K_n=\I$ but also $K_n\mathcal{I}K_n^\dagger\subset\mathcal{I}$ for each $K_n$, i.e. each $K_n$ maps an incoherent state to an incoherent state. With these notions,  Baumgratz \textit{et al.} proposed a  rigorous framework for quantifying coherence, which can be stated as follows  \cite{Baumgratz}.

A functional $C$ can be taken as a coherence measure if it satisfies the four conditions:
\\(C1) $C(\rho)\ge 0$, and $C(\rho)=0$ if and only if $\rho\in\mathcal{I}$;
\\(C2) Monotonicity under incoherent operations, $C(\rho)\ge C(\Lambda(\rho))$ if $\Lambda$ is an incoherent operation;
\\(C3) Monotonicity under selective incoherent operations, $C(\rho)\ge \sum_np_nC(\rho_n)$, where $p_n=\Tr(K_n\rho K_n^\dagger)$, $\rho_n=K_n\rho K_n^\dagger/p_n$, and $\Lambda(\rho)=\sum_nK_n\rho K_n^\dagger$ is an incoherent operation;
\\(C4) Non-increasing under mixing of quantum states, i.e., convexity, $\sum_np_nC(\rho_n)\ge C(\sum_np_n\rho_n)$ for any set of states $\{\rho_n\}$ and any probability distribution $\{p_n\}$.

Based on the rigorous framework, various coherence measures can be constructed. A main family of them are so called convex roof coherence measures, which are defined by extending a functional $C_f$ acting only on pure states to mixed states via the standard convex roof construction \cite{Aberg1,Du1,Yuan,Qi,Liu1,Zhu,Streltsov}. A convex roof coherence measure can be generally expressed as
\begin{equation}
C_f(\rho)=\inf_{\{p_i,\ket{\varphi_i}\}}\sum_ip_iC_f(\ket{\psi_i}),\label{mixed-convex}
\end{equation}
where the infimum is taken over all possible ensembles $\{p_i,\ket{\psi_i}\}$ with $\rho=\sum_ip_i\ket{\psi_i}\bra{\psi_i}$. It is easy to show that $C_f(\rho)$ satisfies conditions (C1)-(C4) for all states $\rho$, as long as $C_f(\ket{\varphi})$ satisfies conditions (C1) and (C3) for all pure states $\ket{\varphi}$ \cite{Yuan}. By following this line, researchers have constructed a number of convex roof coherence measures, including  the coherence of formation \cite{Yuan}, the coherence concurrence \cite{Qi}, the geometric measure of coherence \cite{Streltsov}, the convex roof coherence measure based on fidelity \cite{Liu1}, the convex roof coherence measure based on linear entropy \cite{Peng}, and the convex roof coherence measure based on $\frac{1}{2}$-entropy \cite{Du1}.

\section{\textbf{Theorem on the superadditivity of convex roof coherence measures}}

Since convex roof coherence measures involve an optimization process, it is generally difficult to prove a convex coherence measure superadditive, although it may be easy to prove a convex roof coherence measure non-superadditive. Indeed, a coherence measure can be said non-superadditive  if a counterexample of violating Eq. (\ref{CAB}) is found, but a coherence measure being superadditive means that Eq. (\ref{CAB}) is valid for all states, including all pure and mixed states. The difficulty appears in calculating the coherence of mixed states. We here put forward an approach to  examine the superadditivity of a convex coherence measure, which can steer clear of the difficulty. It can be stated as a theorem.

\emph{Theorem.} A convex roof coherence measure $C_f$  is superadditive for all states if the inequality,
\begin{equation}
C_f(\ket{\varphi}_{AB})\geq C_f\left(\sum_i\sqrt{q_i}\ket{i}_A\right)+\sum_iq_iC_f(\ket{\varphi_i}_B), \label{theorem}
\end{equation}
is satisfied for all pure states $\ket{\varphi}_{AB}=\sum_{ij}c_{ij}\ket{i}_A\ket{j}_B$ with $\sum_{ij}|c_{ij}|^2=1$, where $q_i=\sum_j\abs{c_{ij}}^2$ and $\ket{\varphi_i}_B=\frac{1}{\sqrt{q_i}}\sum_jc_{ij}\ket{j}_B$.

We prove the theorem as follows.

First, we prove that if a coherence measure $C_f$ satisfies Eq. (\ref{theorem}), then the superadditity relation (\ref{CAB}) is fulfilled for all pure state $\ket{\varphi}_{AB}$. To this end, we only need to prove
\begin{equation}
C_f\left(\sum_{i}\sqrt{q_i}\ket{i}_A\right)\geq C_f(\rho_A),\label{theorem-pure1}
\end{equation}
and
\begin{equation}
\sum_iq_iC_f(\ket{\varphi_i}_B)\geq C_f(\rho_B), \label{theorem-pure2}
\end{equation}
where $\rho_A=\tr_B\rho_{AB}=\sum_{ijk}c_{ik}c^*_{jk}\ket{i}_A\bra{j}$ and $\rho_B=\tr_A\rho_{AB}=\sum_iq_i\ket{\varphi_i}_B\bra{\varphi_i}$.

To prove Eq. (\ref{theorem-pure1}), we demonstrate that there exists an incoherent operation that can map $\sum_{i}\sqrt{q_i}\ket{i}_A$ to $\rho_A$. In fact, such an operation can be simply taken as $\Lambda(\cdot)=\sum_{j=1}^{d_B}K_j\cdot K_j^\dag$ with $K_j=\sum_{i=1}^{d_A}\frac{c_{ij}}{\sqrt{q_i}}\ket{i}\bra{i}$. Obviously, the operation defined by $\Lambda$ is incoherent, and it is straightforward to verify that $\Lambda(\sum_{i}\sqrt{q_i}\ket{i}_A)=\rho_A$. Noting that an incoherent operation can never increase the coherence of a state, we then obtain $C_f\left(\sum_{i}\sqrt{q_i}\ket{i}_A\right)\geq C_f(\rho_A)$, i.e., Eq. (\ref{theorem-pure1}).

To prove Eq. (\ref{theorem-pure2}), we use $\rho_B=\sum_ip_i\ket{\psi_i}_B\bra{\psi_i}$ to represent the optimal decomposition of $\rho_B$ that achieves the infimum in Eq. (\ref{mixed-convex}). Since $\rho_B=\sum_iq_i\ket{\varphi_i}_B\bra{\varphi_i}$ is also an ensemble decomposition of $\rho_B$, there must be $\sum_iq_iC_f(\ket{\varphi_i}_B)\geq\sum_ip_iC_f(\ket{\psi_i}_B)\nonumber=C_f(\rho_B)$,  i.e., Eq. (\ref{theorem-pure1}). We then obtain
\begin{equation}
 C(\ket{\varphi}_{AB})\geq C(\rho_A)+C(\rho_B).
\label{CABpure}
\end{equation}

Second, we prove that $C_f$ is superadditive  for all states if it is superadditive for pure states $\rho_{AB}=\ket{\varphi}_{AB}\bra{\varphi}$. To this end, we use $\rho_{AB}=\sum_ip_i\ket{\psi_i}_{AB}\bra{\psi_i}$ to represent one of the optimal decompositions that give $C_f(\rho_{AB})$. By using Eq. (\ref{CABpure}), we have
\begin{eqnarray}
C_f(\rho_{AB})=\sum_ip_iC_f(\ket{\psi_i}_{AB})\geq\sum_ip_i\left(C_f(\rho_i^A)+C_f(\rho_i^B)\right)
\label{theorem-mixed1},
\end{eqnarray}
where $\rho_i^A=\tr_B \ket{\psi_i}_{AB}\bra{\psi_i}$ and $\rho_i^B=\tr_A \ket{\psi_i}_{AB}\bra{\psi_i}$.
According to the convexity of a coherence measure, i.e., condition (C4),  there are $\sum_ip_iC_f(\rho_i^A)\geq C_f(\sum_ip_i\rho_i^A)$ and $\sum_ip_iC_f(\rho_i^B)\geq C_f(\sum_ip_i\rho_i^B)$, which lead to
\begin{eqnarray}
C_f(\rho_{AB})\geq C_f(\sum_ip_i\rho_i^A)+C_f(\sum_ip_i\rho_i^B), \label{theorem-mixed2}.
\end{eqnarray}
Noting that $\rho_A=\sum_ip_i\rho_i^A$ and $\rho_B=\sum_ip_i\rho_i^B$, we finally obtain
\begin{eqnarray}
C_f(\rho_{AB})\geq C_f(\rho_A)+C_f(\rho_B). \label{theorem-mixed}
\end{eqnarray}
This completes the proof of the theorem.

\section{Applications of the theorem}

The above theorem only involves pure states but has nothing to do with mixed states. By verifying the validity of the inequality (\ref{theorem}) for pure states $\ket{\varphi}_{AB}$, one can conclude that Eq. (\ref{CAB}) is valid for all states $\rho_{AB}$, i.e., $C_f$ is of superadditivity. This greatly simplifies the calculations and makes it possible to prove whether a convex roof measure is superadditive. In the following, we will apply our theorem to each of the known convex roof coherence measures to find which of them are superadditive.

\subsection{The coherence of formation}
We show that the coherence of formation is superadditive.

The coherence of formation is defined as
\begin{equation}
C_{for}(\rho)=\inf_{\{p_i, \ket{\varphi_i}\}}\sum_i p_i C_r(\ket{\varphi_i}),\label{cost}
\end{equation}
where $C_r(\ket{\varphi_i})=S(\Delta(\ket{\varphi_i}\bra{\varphi_i})$ with $S(\rho)=-\Tr\rho\log_2\rho$ being the von Neumann entropy. Hereafter, we use $\Delta(\rho)$ to denote the diagonal part of $\rho$, i.e., $\Delta(\rho)=\sum_i\rho_{ii}\ket{i}\bra{i}$. The coherence of formation was first put forward in Ref. \cite{Aberg}, and it was proved to be a coherence measure, i.e., satisfying the conditions (C1-C4), later in Ref. \cite{Yuan}.

To prove the superadditivity of the coherence of formation, we only need to examine the inequality (\ref{theorem}) for pure states $\ket{\varphi}_{AB}=\sum_{ij}c_{ij}\ket{i}_A\ket{j}_B$. Substituting  $\ket{\varphi}_{AB}$ into $C_r(\ket{\varphi}_{AB})=S(\Delta(\ket{\varphi}_{AB}\bra{\varphi}))$, we have
\begin{equation}
C_{for}(\ket{\varphi_{AB}})=-\sum_{ij}\abs{c_{ij}}^2\log_2\abs{c_{ij}}^2. \label{example_11}
\end{equation}
On the other hand, there are
\begin{equation}
C_{for}(\sum_i\sqrt{q_i}\ket{i}_A)=-\sum_iq_i\log_2 q_i \label{example_12},
\end{equation}
\begin{eqnarray}
\sum_iq_iC_{for}(\ket{\varphi_i}_B)=-\sum_iq_i(\sum_j\frac{\abs{c_{ij}}^2}{q_i}\log_2\frac{\abs{c_{ij}}^2}{q_i})=-\sum_{ij}\abs{c_{ij}^2}\log_2\frac{\abs{c_{ij}}^2}{q_i},
\end{eqnarray}
 and therefore
 \begin{eqnarray}
C_{for}(\sum_i\sqrt{q_i}\ket{i}_A)+\sum_iq_iC_{for}(\ket{\varphi_i}_B) &=-\sum_iq_i\log_2 q_i-\sum_{ij}\abs{c_{ij}^2}\log_2\frac{\abs{c_{ij}}^2}{q_i}\nonumber\\
&=-\sum_{ij}\abs{c_{ij}}^2\log_2\abs{c_{ij}}^2 \label{example_13}.
\end{eqnarray}
Comparing Eq. (\ref{example_11}) with Eq. (\ref{example_13}), we immediately obtain $C_{for}(\ket{\varphi}_{AB})= C_{for}\left(\sum_i\sqrt{q_i}\ket{i}_A\right)+\sum_iq_iC_{for}(\ket{\varphi_i}_B)$, which means that Eq. (\ref{theorem}) is fulfilled and therefore the coherence of formation is superadditive.

\subsection{The coherence concurrence}
We show that the coherence concurrence is superadditive.

The  coherence concurrence is defined as
\begin{equation}
C_C(\rho)=\inf_{\{p_i, \ket{\varphi_i}\}}\sum_i p_i C_{l_1}(\ket{\varphi_i}),\label{concurrence}
\end{equation}
where $C_{l_1}(\rho)=\sum_{i\neq j}\abs{\rho_{ij}}$ is the $l_1$ norm of coherence \cite{Baumgratz}. The  coherence concurrence was first put forward in Ref. \cite{Du1}, and rigourously proved in Ref. \cite{Qi}.

To prove the superadditivity of the  coherence concurrence, we calculate $C_C(\ket{\varphi}_{AB})$ with    $\ket{\varphi}_{AB}=\sum_{ij}c_{ij}\ket{i}_A\ket{j}_B$, and have
\begin{eqnarray}
C_C(\ket{\varphi}_{AB})&=\left(\sum_{i,k}\abs{c_{ik}}\right)^2-1. \label{example21}
\end{eqnarray}
On the other hand, there are
\begin{eqnarray}
C_C(\sum_i\sqrt{q_i}\ket{i}_A)=\sum_{i\neq j}\sqrt{\sum_{k,l}\abs{c_{ik}c_{jl}}^2},
\end{eqnarray}
\begin{eqnarray}
\sum_iq_iC_C(\ket{\varphi_i}_B)=\sum_{i,k,l}\abs{c_{ik}c_{il}}-1,
\end{eqnarray}
and therefore
\begin{eqnarray}
C_C(\sum_i\sqrt{q_i}\ket{i}_A)+\sum_iq_iC_C(\ket{\varphi_i}_B)
&=\sum_{i\neq j}\sqrt{\sum_{k,l}\abs{c_{ik}c_{jl}}^2}+\sum_{i,k,l}\abs{c_{ik}c_{il}}-1\nonumber\\
&\leq \sum_{i\neq j}\sum_{k,l}\abs{c_{ik}c_{jl}}+\sum_{i,k,l}\abs{c_{ik}c_{il}}-1\nonumber\\
&=\sum_{i,j,k,l}\abs{c_{ik}c_{jl}}-1.\label{example23}
\end{eqnarray}
Comparing Eq. (\ref{example21}) with Eq. (\ref{example23}), we immediately obtain $C_C(\ket{\varphi}_{AB})\geq C_C\left(\sum_i\sqrt{q_i}\ket{i}_A\right)+\sum_iq_iC_C(\ket{\varphi_i}_B)$, which means that Eq. (\ref{theorem}) is fulfilled and therefore the coherence concurrence is superadditive.

\subsection{The geometric measure of coherence} \label{subsection3}
We show that the geometric measure of coherence is non-superadditive.

The geometric measure of coherence is defined as
\begin{eqnarray}
C_g({\rho})=\inf_{\{p_i,\ket{\varphi_i}\}}\sum_i p_i\left(1- F(\ket{\varphi_i}, \delta)\right) , \label{fidelity}
\end{eqnarray}
where $F(\rho,\delta)=\left(\Tr(\sqrt{\sqrt{\rho}\delta\sqrt{\rho}})\right)^2$ is the Uhlmann fidelity \cite{Uhlmann}. This measure was put forward in Ref. \cite{Streltsov}. There is $C_g(\ket{\varphi})=1-\abs{c_i}^2_{\max}$ for pure states $\ket{\varphi}=\sum_ic_i\ket{i}$ \cite{Streltsov1}.

To prove the geometric measure of coherence non-superadditive, we give a counterexample to inequality ({\ref{theorem}}).
The counterexample can be taken as $\ket{\varphi}_{AB}=\frac{1}{2}(\ket{11}+\ket{12}+\ket{21}+\ket{22})$. For this state, we have
$C_g(\ket{\varphi}_{AB})=\frac{3}{4}$, $C_g(\sum_i\sqrt{q_i}\ket{i}_A)=C_g(\frac{1}{\sqrt{2}}\ket{1}_A+\frac{1}{\sqrt{2}}\ket{2}_A)=\frac{1}{2}$, and  $\sum_iq_iC_g(\ket{\varphi_i}_B)=C_g(\frac{1}{\sqrt{2}}\ket{1}_B+\frac{1}{\sqrt{2}}\ket{2}_B)=\frac{1}{2}$. Then, there is
$C_g(\ket{\varphi}_{AB})=\frac{3}{4}< C_g\left(\sum_i\sqrt{q_i}\ket{i}_A\right)+\sum_iq_iC_g(\ket{\varphi_i}_B)=1$, which violates the condition in the theorem. In this case, it is suspected that the geometric measure of coherence is non-superadditive. However, its non-superadditivity cannot be decided only by the violation of the inequality (\ref{theorem}), since the inequality in our theorem is only a sufficient condition of superadditivity. To confirm the non-superadditivity of $C_g$,  we use the definition relation of superadditivity, i.e., Eq. (\ref{CAB}). In fact, since $\ket{\varphi}_{AB}=\frac{1}{2}(\ket{11}+\ket{12}+\ket{21}+\ket{22})$ is a separable state, there are always  $C_g\left(\sum_i\sqrt{q_i}\ket{i}_A\right)=C_g(\rho_A)$ and $\sum_iq_iC_g(\ket{\varphi_i}_B)=C_g(\rho_B)$, and therefore Eq. (\ref{CAB}) is not valid, too.

\subsection{Convex roof coherence measure based on fidelity and that based on linear entropy}
We show that both convex roof coherence measure based on fidelity and convex roof coherence measure based on linear entropy are non-superadditive, too.

Convex roof coherence measure based on fidelity is defined as
\begin{equation}
C_F(\rho)=\inf_{\{p_i,\ket{\varphi_i}\}}\sum_ip_i\sqrt{1-F(\ket{\varphi_i},\delta)},\label{fide_liu}
\end{equation}
where $F(\rho,\delta)=\left(\Tr(\sqrt{\sqrt{\rho}\delta\sqrt{\rho}})\right)^2$ is the Uhlmann fidelity. It was put forward in Ref. \cite{Liu1}.

Convex roof coherence measure based on linear entropy is defined as
\begin{equation}
C_L(\rho)=\inf_{\{p_i, \ket{\varphi_i}\}}\sum_i p_i C_L(\ket{\varphi_i}),\label{linentropy}
\end{equation}
where $C_L(\ket{\varphi})=\sum_i\abs{c_i}^4$ for $\ket{\varphi}=\sum_ic_i\ket{i}$. It was put forward in Ref. \cite{Peng}.

To prove the convex roof coherence measure based on fidelity non-superadditive, we take the same state $\ket{\varphi}_{AB}=\frac{1}{2}(\ket{11}+\ket{12}+\ket{21}+\ket{22})$, as done in Subsection \ref{subsection3}. There are
$C_F(\ket{\varphi_{AB}})=\frac{\sqrt{3}}{2}$, $C_F(\sum_i\sqrt{q_i}\ket{i}_A)=\frac{1}{\sqrt{2}}$, and  $\sum_iq_iC_F(\ket{\varphi_i}_B)=\frac{1}{\sqrt{2}}$, which does not fulfill Eq. (\ref{theorem}) as well as Eq. (\ref{CAB}).
Similarly, to prove the convex roof coherence measure based on linear entropy non-superadditive, we again take $\ket{\varphi}_{AB}=\frac{1}{2}(\ket{11}+\ket{12}+\ket{21}+\ket{22})$. There are $C_L(\ket{\varphi}_{AB})=\frac{1}{4}$, $C_F(\sum_i\sqrt{q_i}\ket{i}_A)=\frac{1}{2}$, and $\sum_iq_iC_L(\ket{\varphi_i}_B)=\frac{1}{2}$, which does not fulfill Eq. (\ref{theorem}) as well as Eq. (\ref{CAB}), too. Hence, both the convex roof coherence measure based on fidelity and that based on linear entropy are non-superadditive.

\subsection{The convex roof coherence measure based on $\frac12$-entropy}
We show that the  convex roof coherence measure based on $\frac12$-entropy is non-superadditive.

Convex roof coherence measure based on $\frac12$-entropy is defined as
\begin{equation}
C_{\frac12}(\rho)=\inf_{\{p_i, \ket{\varphi_i}\}}\sum_i p_i C_{\frac12}(\ket{\varphi_i}),\label{alp_entropy}
\end{equation}
where $C_{\frac12}(\ket{\varphi})=2\log_2(\sum_{i=1}^d\abs{c_i})$ for $\ket{\varphi}=\sum_ic_i\ket{i}$. It was proposed in Ref. \cite{Du1}.
To show that this measure does not fulfill the inequality (\ref{theorem}), a counterexample can be taken as $\ket{\varphi}_{AB}=\sqrt{\frac{53}{100}}\ket{11}+\sqrt{\frac{11}{50}}\ket{12}+\sqrt{\frac{11}{50}}\ket{21}+\sqrt{\frac{3}{100}}\ket{22}$.
For this state, there are  $C_{\frac12}(\ket{\varphi}_{AB})=2\log_2\frac{\sqrt{53}+\sqrt{3}+2\sqrt{22}}{10}$,   $C_{\frac12}(\sum_i\sqrt{q_i}\ket{i}_A)=2\log_2\frac{1+\sqrt{3}}{2}$, and
$\sum_iq_iC_{\frac12}(\ket{\varphi_i}_B)=\frac{3}{2}\log_2\frac{\sqrt{53}+\sqrt{22}}{5\sqrt{3}}+\frac{1}{2}\log_2\frac{\sqrt{22}+\sqrt{3}}{5}$. We then have
$C_{\frac12}(\ket{\varphi}_{AB})< C_{\frac12}\left(\sum_i\sqrt{q_i}\ket{i}_A\right)+\sum_iq_iC_{\frac12}(\ket{\varphi_i}_B)$, which means that $C_{\frac12}$ does not fulfill Eq. (\ref{theorem}).

To confirm that the  convex roof coherence measure based on $\frac12$-entropy is non-superadditive. We need to examine Eq. (\ref{CAB}) with $\rho_{AB}=\ket{\varphi}_{AB}\bra{\varphi}$.  By following the same method used for obtaining $C_F(\rho)$ in Ref. \cite{Liu1}, we can obtain the expression of  $C_{\frac12}(\rho)$ for single-qubit states $\rho$,
\begin{equation}
 C_{\frac12}(\rho)=2\log_2\left(\sqrt{\frac{1+\sqrt{1-C_{l_1}(\rho)^2}}{2}}+\sqrt{\frac{1-\sqrt{1-C_{l_1}(\rho)^2}}{2}}\right),\label{ex_alp_entropy}
\end{equation}
where $C_{l_1}(\rho)$ is the $l_1$ norm of coherence.  With the aid of Eq. (\ref{ex_alp_entropy}), it is easy to work out  $C_{\frac12}(\ket{\varphi}_{AB})-C_{\frac12}(\rho_A)-C_{\frac12}(\rho_B)=-0.0096$ with $\rho_A=\Tr_B\ket{\varphi}_{AB}\bra{\varphi}$ and $\rho_B=\Tr_A\ket{\varphi}_{AB}\bra{\varphi}$. This indicates that $C_{\frac12}$ is non-superadditive.

\section{Remarks and Conclusions}
Quantifying coherence has received increasing attention, and considerable work has been directed towards finding links between coherence measures and quantum correlations. Superadditivity of a coherence measure describes the trade-off relations between the coherence of a bipartite system and that of its subsystems and it is a precondition of defining a discordlike correlation based on the coherence measure. In this paper, we have put forward a theorem on the superadditivity of convex roof coherence measures, which provides a sufficient condition to identify the convex roof coherence measures fulfilling the superadditivity. By applying our theorem to each of the known convex roof coherence measures, we prove that the coherence of formation and the coherence concurrence are superadditive, while the geometric measure of coherence, the convex roof coherence measure based on linear entropy, the convex roof coherence measure based on fidelity, and convex roof coherence measure based on $\frac{1}{2}$-entropy are non-superadditive. Noting that some distance-based coherence measures have been used to define a discordlike correlation \cite{Guo,Tan,Wang}, our results indicate that a discordlike correlation of the form $I_{c}(\rho_{AB})=C(\rho_{AB})-C(\rho_A)-C(\rho_B)$ can be defined based on the convex roof coherence measures with the superadditivity, such as the coherence of formation and the coherence concurrence.

In passing, we would like to point that the expression of the sufficient condition  in our theorem is not unique. In stead of Eq. (\ref{theorem}), an alternative expression of the sufficient condition can be taken as
\begin{equation}
C_f(\ket{\varphi}_{AB})\geq \sum_jp_jC_f(\ket{\varphi_j}_A)+\sum_iq_iC_f(\ket{\varphi_i}_B),\label{theorem2}
\end{equation}
where $p_j=\sum_i\abs{c_{ij}}^2$, $\ket{\varphi_j}_A=\frac{1}{\sqrt{p_j}}\sum_ic_{ij}\ket{i}_A$, and all the others are the same as in the theorem. Compared with Eq. (\ref{theorem}), Eq. (\ref{theorem2}) is more accuracy in the sense that the right hand side of Eq. (\ref{theorem2}) is smaller than that of Eq. (\ref{theorem}), but Eq. (\ref{theorem}) is more convenient to use.

\ack
C.L.L. acknowledges support from the National Natural Science Foundation of China through Grant No. 11575101. Q.M.D. acknowledges support from the National Natural Science Foundation of China through Grant No. 11775129. D.M.T. acknowledges support from the National Basic Research Program of China through Grant No. 2015CB921004.

\end{document}